\documentclass[pra,twocolumn,preprintnumbers,amsmath,amssymb,nofootinbib,floatfix]{revtex4}

\usepackage{graphicx, bm, hyperref, color}

\makeatletter

\def\graphicscale{\twocolumn@sw{0.3}{0.4}}
\def\graphicthreescale{\twocolumn@sw{0.3}{0.4}}

\begin{document}

\title{Ground-state fidelity at first-order quantum transitions}

\author{Davide Rossini}
\affiliation{Dipartimento di Fisica dell'Universit\`a di Pisa
        and INFN, Largo Pontecorvo 3, I-56127 Pisa, Italy}

\author{Ettore Vicari} 
\affiliation{Dipartimento di Fisica dell'Universit\`a di Pisa
        and INFN, Largo Pontecorvo 3, I-56127 Pisa, Italy}

\date{\today}

\begin{abstract}
We analyze the scaling behavior of the fidelity, and the corresponding
susceptibility, emerging in finite-size many-body systems whenever a
given control parameter $\lambda$ is varied across a quantum phase
transition.  For this purpose we consider a finite-size scaling (FSS)
framework.  Our working hypothesis is based on a scaling assumption of
the fidelity in terms of the FSS variables associated to $\lambda$ and
to its variation $\delta \lambda$.  This framework entails the FSS
predictions for continuous transitions, and meanwhile enables to
extend them to first-order transitions, where the FSS becomes
qualitatively different.  The latter is supported by analytical and
numerical analyses of the quantum Ising chain along its first-order
quantum transition line, driven by an external longitudinal field.
\end{abstract}

\maketitle


\section{Introduction}

Quantum transitions (QTs) in many-body systems are related to
significant changes of the ground state and low-excitation properties,
induced by small variations of a driving
parameter~\cite{Sachdev-book, SGCS-97}.  They are continuous when the
ground state of the system changes continuously at the transition
point, and correlation functions develop a divergent length
scale. Instead, they are of first order when the ground-state
properties are discontinuous across the transition point, generally
arising from level crossings in the infinite-volume limit.
In view of their key role played in several contexts of modern
statistical mechanics, quantum information and condensed matter
physics, it is of crucial importance to devise suitable tools for a
proper characterization of their main features.  To this purpose,
different quantum-information based concepts have been recently put
forward, in order to spotlight ground-state variations at QTs, such as
the entanglement, as well as the fidelity and its
susceptibility~\cite{AFOV-08, Gu-10, BAB-17}.  The net advantage of
these approaches is that they do not rely on the identification of an
order parameter with the corresponding symmetry breaking pattern.

In particular, the fidelity quantifies the overlap between the ground
states of quantum systems sharing the same Hamiltonian, but associated
with different Hamiltonian parameters~\cite{AFOV-08, Gu-10, BAB-17}.
The concept of the fidelity and, more generally, of the geometric
tensor has recently gained considerable attraction, in the field of
quantum information and computation.  The reason is related to its
fundamental importance as a basic tool to analyze the variations of a
given quantum state in the Hilbert space.  The usefulness of the
fidelity as a tool to distinguish quantum states can be traced back to
Anderson's orthogonality catastrophe~\cite{Anderson-67}: the overlap
of two many-body ground states corresponding to Hamiltonians differing
by a small perturbation vanishes in the thermodynamic limit.  It is
thus tempting to quantify how this paradigm gets realized in many-body
systems at QTs, where significantly different behaviors are expected
with respect to systems in normal conditions.  Besides that, the
fidelity susceptibility covers a central role in quantum estimation
theory~\cite{BC-94, Paris-09}, being proportional to the Fisher
information.  The latter indeed quantifies the inverse of the smallest
variance in the estimation of the varying parameter, such that, in
proximity of QTs, metrological performances are believed to
drastically improve~\cite{ZPC-08, IKCP-08}.

The last decade has seen the birth of an intense theoretical activity
focusing on the behavior of the fidelity and of the corresponding
susceptibility (more generally, of the geometric tensor)~\cite{ZP-06,
  YLG-07, VZ-07} at continuous QTs (CQTs).  In quantum many-body
systems, the establishment of a non-analytic behavior has been
exploited to evidence CQTs in several different contexts, which have
been deeply scrutinized both analytically and numerically.  We quote,
for example, free-fermion models~\cite{CGZ-07, MPD-11, RD-11,
  Damski-13, LZW-18}, interacting spin~\cite{CWHW-08, SAC-09, LLZ-09,
  AASC-10, Sirker-10, Nishiyama-13, SKV-14} and particle
models~\cite{BV-07, MHCFR-11, CMR-13, WLIMT-15, HWWW-16, Kettemann-16,
  SS-17}, as well as systems presenting peculiar
topological~\cite{YGSL-08, OS-14, KLS-16} and non-equilibrium
steady-state transitions~\cite{BGZ-14, MP-17}. However a
characterization of first-order QTs (FOQTs) in this context is still
missing, despite the fact that they are of great phenomenological
interest.  Indeed they occur in a large variety of many-body systems,
including quantum Hall samples~\cite{PPBWJ-99}, itinerant
ferromagnets~\cite{VBKN-99}, heavy fermion metals~\cite{UPH-04,
  Pfleiderer-05, KRLF-09}, disordered systems~\cite{JKKM-08, JLSZ-10}
and infinite-range models~\cite{YKS-10, SN-12}.

We also stress that, to achieve a deep understanding of QTs from
the outcomes of numerical simulations or of quantum-simulation
experiments, it is fundamental to exploit the impact of having a
finite size.  The natural theoretical context where to set up the
analysis is the finite-size scaling (FSS) framework, that has been
proven to be effective in proximity of any type of QTs. Indeed the
emergence of FSS limits has been predicted both for
CQTs~\cite{Sachdev-book, SGCS-97, CPV-14} and for FOQTs~\cite{CNPV-14}, 
as well as to describe the quantum dynamics of
finite-size many-body systems subject to time-dependent
perturbations~\cite{PRV-18, PRV-18b}.  This formalism has been
successfully applied in a variety of systems, for observables such
as the free energy, the energy gaps of the first low-lying levels,
correlation functions, as well as in the presence of different
boundary conditions~\cite{CPV-15, CPV-15b, PRV-18c}.  Recently, it
has been also used to study quantum-information based concepts, such
as entanglement~\cite{AFOV-08, CPV-14, CCD-09} and other indicators
of quantum correlations~\cite{DS-18}.  Some results for the FSS of
the fidelity have been obtained in specific situations at CQTs, such
as for the quantum Ising chain in a transverse field (see the
discussion in Sec.~\ref{sec:CQT}), and by means of quite complicated
methods~\cite{Gu-10}.

In this paper we present a unified picture for the scaling
behavior of the fidelity, and its susceptibility, emerging in
many-body systems whenever a given control parameter is varied across
any type of QT.  Since ground-state overlaps related to
variations of the Hamiltonian parameters are naturally defined only
for finite quantum systems, whose ground-state wave functions are
normalizable, we consider finite-size systems and focus on the
asymptotic large-volume behavior of the fidelity, defined in the limit
of small variations of the parameter driving the QT.  The FSS
theory constitutes the optimal framework to discuss this issue. It
turns out to be especially effective to provide the power or
exponential laws describing the size dependence of fidelity and its
susceptibility when the system is driven across a QT. In particular
we discuss FOQTs for the first time.

Assuming that the fidelity of finite systems is an analytic function
of the relevant scaling variables associated to the driving parameter
and to its variation, we put forward a FSS behavior that entails the
expected power-law divergences associated with CQTs, meanwhile
enabling to extend the analysis to FOQTs. In the latter, the type of
divergence is controlled by the closure of the gap between the two
lowest energy levels, being exponential in most of the cases.  A
scaling theory for the fidelity provides a simple and intuitive route
towards a complete understanding of the behaviors of finite-size
many-body systems at CQTs and FOQTs, which is mandatory to distinguish
them, and obtain correct interpretations of experimental and numerical
results at QTs.

The paper is organized as follows.  In Sec.~\ref{sec:FSS_fid}, we
discuss the theory underlying our FSS framework for the fidelity and
its susceptibility, holding whenever a many-body system undergoes a
QT.  Our predictions are then verified in Sec.~\ref{sec:Ising} for the
paradigmatic quantum Ising model driven by an additional external
longitudinal field, exhibiting a rich phase diagram.  In this context,
we focus on both CQTs (\ref{sec:CQT}) and FOQTs (\ref{sec:FOQT}) with
different boundary conditions.  A summary of our results, together
with the perspectives, is finally drawn in Sec.~\ref{sec:concl}.

\section{Finite-size scaling  of the fidelity and its susceptibility}
\label{sec:FSS_fid}

\subsection{The fidelity and its susceptibility}
\label{subfidesusc}

We define our setting by considering a $d$-dimensional quantum
many-body system of size $L^d$, with Hamiltonian
\begin{equation}
H(\lambda) = H_0 + \lambda H_I ,
\label{hdef}
\end{equation}
where $[H_0,H_I]\neq 0$ and the parameter $\lambda$ drives the QT
located at $\lambda=0$. The fidelity
\begin{equation}
F(\lambda,\delta\lambda,L) \equiv | \langle
\Psi_0(\lambda+\delta\lambda,L) | \Psi_0(\lambda,L) \rangle|
\label{fiddef}
\end{equation}
is a geometrical object that can be used to monitor the changes of the
ground-state wave function $|\Psi_0(\lambda,L)\rangle$ when varying
the control parameter $\lambda$ by a small amount $\delta \lambda$
around its transition value.  Assuming $\delta\lambda$ sufficiently
small, one can expand Eq.~\eqref{fiddef} in powers of
$\delta\lambda$:~\cite{Gu-10}
\begin{equation}
  F(\lambda,\delta\lambda,L) 
  = 1 - \tfrac12 (\delta\lambda^2) \, 
\chi_F(\lambda,L) + O(\delta\lambda^3),
  \label{expfide}
\end{equation}
where $\chi_F$ defines the fidelity susceptibility.  The cancellation
of the linear term of the expansion is essentially related to the fact
that the fidelity is bounded, i.e., $F\le 1$.  Standard perturbation
theory allows us to also write $\chi_F$ as~\cite{Gu-10}:
\begin{equation}
  \chi_F(\lambda,L) = \sum_{n>0}
{|\langle \Psi_n(\lambda,L)| H_I |\Psi_0(\lambda,L)\rangle |^2 \over
[E_n(\lambda,L) -
E_0(\lambda,L)]^2},  \label{chifpert} 
\end{equation}
where $|\Psi_n(\lambda,L)\rangle$ is the Hamiltonian eigenstate
corresponding to the eigenvalue $E_n(\lambda,L)$ (notice that 
the index $n=0$ labels ground-state quantities).

As we shall see below, the interplay between $\lambda$ and $L$ 
at QTs can be suitably described within FSS frameworks 
at both CQTs~\cite{SGCS-97,CPV-14} and FOQTs~\cite{CNPV-14}.

\subsection{Finite-size scaling  at  continuous quantum transitions}
\label{fsscont}

Singular behaviors at QTs are observed in the infinite-volume limit.
If the size $L$ of the system is finite, all properties are generally
analytic as a function of the quantity driving the
transition. However, around the transition point, low-energy
thermodynamic quantities and large-scale structural properties undergo
peculiar FSS behaviors depending only on the nature and on the general
properties of the transition.  Understanding these finite-size
properties is of primary importance for a correct and unambiguous
interpretation of experimental or numerical data when phase
transitions are investigated in relatively small systems (see, e.g.,
Refs.~\cite{Barber-83, Privman-90, PV-02, GKMD-08}) or in particle
systems trapped by external forces, as in cold-atom experiments (see,
e.g., Ref.~\cite{BDZ-08}).

The modern theory of FSS delineates the standard roadmap to
investigate these issues at phase transitions.  It was originally
developed in the context of critical phenomena, and formulated in the
classical framework~\cite{FB-72,Barber-83}.  At continuous
transitions, FSS is observed when the length scale $\xi$ of the
critical modes becomes comparable with $L$.  For large values of $L$,
this regime presents universal features, shared by all systems whose
transition belongs to the same universality class.  Analogous
behaviors emerge at CQTs~\cite{SGCS-97,CPV-14}, where the FSS
framework allows one to characterize the finite-size dependence of the
low-energy properties of quantum many-body systems, in particular the
low-excitation spectrum, the correlation functions, etc.  The critical
behavior is generally characterized by power laws, with universal
exponents determined by the universality class of the CQT. They do not
depend on the microscopic details of the quantum model, but only on
some global properties, such as the spatial dimension, the symmetry,
the nature of the interactions (whether they are short-range or
long-range). In particular, relevant universal exponents are the
renormalization-group (RG) dimension $y_\lambda$ of the parameter
$\lambda$ driving the transition, and the dynamic exponent $z$
associated with the scaling behavior of the gap, i.e., the energy
difference of the lowest states~\cite{Sachdev-book}.

The FSS limit is generally obtained at large $L$, keeping an
appropriate combination $\kappa$ of $\lambda$ and $L$ fixed. At CQTs,
this is generally given by~\cite{CPV-14}
\begin{equation}
\kappa = \lambda \, L^{y_\lambda} .
\label{lappaCQT}
\end{equation}
Generic observables $O$ behave as~\cite{SGCS-97,CPV-14}
\begin{equation}
O(\lambda,L)\approx L^{-y_o} f_O(\kappa),
\label{fssex}
\end{equation}
where $y_o$ is the RG dimension associated with $O$, and $f_O(\kappa)$
a scaling function.  Note that the universal power laws at CQTs do not
depend on the boundary conditions, which only affect the scaling
functions.

The temperature $T$ gives rise to an additional relevant perturbation
at CQTs.  Within the FSS framework, it is taken into account by adding
a further dependence of the scaling functions on the scaling
variable~\cite{SGCS-97, Sachdev-book}
\begin{equation}
\tau \sim T/\Delta_0(L),\qquad \Delta_0(L)\sim L^{-z},
\label{taudef}
\end{equation}
where $\Delta_0(L)$ is the energy difference of the
lowest states at the transition point of CQT and $z$ is the dynamic
exponent.

We are now in the position to discuss the scaling behavior of the
fidelity $F(\lambda,\delta\lambda,L)$ and its susceptibility
$\chi_F(\lambda,L)$, assuming that both $\lambda$ and
$\lambda+\delta\lambda$ are sufficiently small to be in the transition
region.  We conjecture that the zero-temperature scaling is given by
\begin{equation}
F(\lambda,\delta\lambda,L) \approx {\cal F}(\kappa,\delta\kappa) ,
\label{fisca}
\end{equation}
where $\delta\kappa$ is the variation of $\kappa$ corresponding to
$\delta\lambda$. The scaling relation~\eqref{fisca} is quite natural,
noting that $F(\lambda,0,L)=1$ and that a regular expansion around
$\delta\lambda=0$ is expected at finite volume. Correspondingly, we
expect ${\cal F}(\kappa,0)=1$ and a regular behavior around
$\delta\kappa=0$.  The FSS of $\chi_F$ can be immediately derived from
Eq.~(\ref{fisca}), by expanding ${\cal F}$ in powers of $\delta
\kappa$,
\begin{equation}
  {\cal F}(\kappa,\delta\kappa) = 1 - \tfrac12 (\delta\kappa^2) \,
  {\cal F}_2(\kappa) + O(\delta\kappa^3), 
  \label{fkexp}
\end{equation}  
and matching it with Eq.~(\ref{expfide}):
\begin{equation}
\chi_F(\lambda,L) \approx (\delta\kappa/\delta\lambda)^2 \, {\cal
  F}_2(\kappa).
\label{chifscal}
\end{equation}
This implies
\begin{equation}
\chi_F(\lambda,L)
 \approx L^{2 y_\lambda} {\cal F}_2(\kappa).
\label{cqtchil}
\end{equation}

We stress that this obtained FSS power law perfectly agrees with
earlier (apparently more involved) derivations, which have been
obtained by means of alternative scaling
arguments~\cite{Gu-10, footnote1}.  However, an important feature of
our novel derivation is that the validity of Eq.~(\ref{chifscal}) can
be extended to FOQTs as well, by inserting the appropriate scaling
variable $\kappa$, see below.  In such case, for transitions
based on the avoided crossing of two levels, the
conjecture~\eqref{fisca} can be straightforwardly justified by means
of a simple calculation on the effective Hamiltonian, as well (see
App.~\ref{app:2lev}).

\subsection{Finite-size scaling  at first-order quantum transitions}
\label{fssfo}

FSS behaviors also develop at FOQTs, although with significant
differences~\cite{CNPV-14}. In particular, they turn out to be
more sensitive to the boundary conditions, which may give rise to
different functional dependencies of the corresponding scaling
variable $\kappa$, leading to both exponential and power laws.

FOQTs generally arise from level crossings. However level crossings
can only occur in the infinite-volume limit (in the absence of
particular conservation laws).  In a finite system, the presence of a
nonvanishing matrix element among these states lifts the degeneracy,
giving rise to the phenomenon of avoided level crossing.  Here the FSS
is controlled by the energy difference $\Delta(\lambda,L)$ of the
avoiding levels, in particular by
\begin{equation}
\Delta_0(L) \equiv \Delta(\lambda=0,L). 
\label{delta0def}
\end{equation}
The appropriate FSS variable is generally given by~\cite{CNPV-14}
\begin{equation}
\kappa \sim {E_\lambda(\lambda,L) \over \Delta_{0}(L)} ,
\label{kappafoqt}
\end{equation}
$E_\lambda$ being the energy variation associated with the $\lambda$
term (we assume $E_\lambda= 0$ at the transition point).  The FSS
limit is defined by the large-$L$ limit, keeping $\kappa$ fixed.
However, it is important to remark that the FSS at FOQTs is more
complex than that at CQTs, because it may significantly depend on the
boundary conditions~\cite{CNPV-14, CPV-15, CPV-15b, PRV-18c}:
the gap $\Delta_0(L)$ may depend on the size $L$ either exponentially
(as it occurs in typical situations), or even as a power law.  As a
matter of fact, the FOQT scenario based on the avoided crossing of two
levels is not always realized, depending on the boundary
conditions (see below); indeed, in some cases the energy difference
$\Delta_0(L)$ of the lowest levels may even show a power-law
dependence on $L$.  However, as we shall see, the scaling variables
$\kappa$ obtained using the corresponding $\Delta_0(L)$ turn out to be
appropriate as well.

In order to derive the scaling behavior of the fidelity and its
susceptibility, cf. Eqs.~(\ref{fiddef}) and (\ref{expfide}), we can
repeat the scaling arguments of Sec.~\ref{fsscont} done at CQTs.
Therefore, assuming again that both $\lambda$ and
$\lambda+\delta\lambda$ are sufficiently small to be in the transition
region, we obtain Eqs.~(\ref{fisca}) and (\ref{chifscal}) as well, but
with the appropriate scaling variable $\kappa$ given now by
Eq.~(\ref{kappafoqt}).  In particular, for FOQTs we obtain
\begin{equation}
\chi_F(\lambda,L) \approx \Delta_0(L)^{-2} \,
(\partial E_\lambda/\partial\lambda)^2 \, {\cal F}_2(\kappa).
\label{chiflal}
\end{equation}

We note that at FOQTs the finite-size dependence of the fidelity
susceptibility appears to be closely connected with the size
dependence of the energy difference of the lowest levels.  Since the
gap can be exponentially suppressed for some types of boundary
conditions, such as periodic or equal and fixed boundary conditions,
for which $\Delta_0(L)\sim e^{-a L^d}$, in such cases we expect
corresponding exponentially large behaviors for the fidelity
susceptibility, $\chi_F \sim e^{c L^d}$ at the transition point
(Sec.~\ref{sec:PBC-OBC} and~\ref{sec:EFBC} ).  For other types of
boundary conditions, such as antiperiodic boundary conditions, for
which $\Delta_0(L) \sim L^{-b}$, we expect a power-law behavior of the
fidelity susceptibility with $L$ (Sec.~\ref{sec:ABC}), as it happens
in proximity of CQTs.

\subsection{Finite-size scaling  at finite temperature}
\label{fssfite}

The above FSS framework, both for CQTs and for FOQTs, can be
generalized to a finite temperature $T$, as well~\cite{Gu-10}.  In
such case, the quantum system is described by the density matrix
\begin{equation} 
\rho_\lambda\equiv \rho(\lambda,T,L) = Z^{-1} \sum_n e^{-E_n/ k_B T} |
\Psi_n\rangle \langle \Psi_n |,
\label{rhodef}
\end{equation}
where $Z= \sum_n \langle \Psi_n | e^{-E_n / k_B T} | \Psi_n\rangle$
denotes the partition function.  The fidelity between two mixed states
can be defined as~\cite{Uhlmann-76}:
\begin{equation}
F(\lambda,\delta\lambda,T,L) = {\rm Tr} \sqrt{\sqrt{\rho_\lambda} \,
  \rho_{\lambda+\delta\lambda} \, \sqrt{\rho_\lambda}},
\label{fiddefT}
\end{equation}
which reduces to Eq.~\eqref{fiddef} for $T\to 0$.  The corresponding
fidelity susceptibility can be extracted analogously to
Eq.~(\ref{expfide}).  At a QT, the $T=0$ scaling (\ref{fisca}) can be
straightforwardly extended to keep into account the temperature, by
adding a further scaling variable $\tau = T/\Delta_0(L)$, so that
\begin{equation}
F(\lambda,\delta\lambda,T,L) \approx {\cal
  F}(\kappa,\delta\kappa,\tau).
\label{fiscaT}
\end{equation}
This scaling equation holds at both CQTs and FOQTs, with the
appropriate definitions of scaling variables.  In particular, $\tau =
T/\Delta_0(L) \sim T L^{z}$ at CQTs, where $z$ is the dynamic
exponent.

\section{Results for the quantum Ising chain}
\label{sec:Ising}

We now verify the above general FSS predictions by presenting
analytical and numerical evidence for the paradigmatic one-dimensional
quantum Ising model in the presence of transverse and longitudinal
fields.  Its Hamiltonian reads
\begin{equation}
H_{\rm Is} = - J \, \sum_{\langle i,j \rangle}
\sigma^{(3)}_i \sigma^{(3)}_j - g\, \sum_i
\sigma^{(1)}_i - h \,\sum_i \sigma^{(3)}_i ,
\label{hedef}
\end{equation}
where $\sigma^{(k)}$ are the Pauli matrices, the first sum is over all
bonds connecting nearest-neighbor sites $\langle i,j \rangle$,
while the other sums are over the $L$ sites.  Hereafter we assume
$\hslash=k_B=1$, $J=1$ and $g>0$.

At $g=1$ and $h=0$, the model undergoes a CQT belonging to the
two-dimensional Ising universality class, separating a disordered
phase ($g>1$) from an ordered ($g<1$) one~\cite{Sachdev-book}. For any
$g<1$, the field $h$ drives FOQTs along the $h=0$ line.  Relevant
observables at the FOQT line are the energy difference $\Delta(h,L)$
of the lowest levels and the magnetization $m = L^{-1} \langle
\sum_i \sigma^{(3)}_i \rangle$.  In the following, we
are interested in the behavior of the ground-state
fidelity~\eqref{fiddef} arising from changes of the longitudinal field
$h \equiv \lambda$, keeping $g$ fixed.  The fidelity susceptibility is
obtained by expanding $F$ to second order in powers of $\delta h$.

\subsection{FSS at the continuous transition}
\label{sec:CQT}

At the CQT, located at $g=1$, $h=0$, the system is expected to
develop the asymptotic FSS behavior in Eq.~(\ref{fisca}).  Let us
analyze two situations in which the control parameter is assumed to
be either $h$ or $g$, and it is tuned through the CQT point.

We first consider the case in which the longitudinal field $h$ is
varied across the value $h=0$, while the transverse field strength
is kept fixed at $g=1$. The exponent $y_h$ entering the
corresponding scaling variable $\kappa = h L^{y_h}$ [see
  Eq.~\eqref{lappaCQT}] is provided by the RG dimension of the
longitudinal magnetic field $h$, i.e.,
\begin{equation}
  y_h = (d + z  + 2 - \eta)/2.
\end{equation}
For the quantum Ising ring in Eq.~\eqref{hedef}, we have $d=1$,
$z=1$ and $\eta=1/4$, thus $y_h = 15/8$.  Details on the
derivation of the Ising critical exponents and of the 
RG dimension $y_h$ are provided, e.g., in Ref.~\cite{CPV-14}.
Correspondingly, inserting such value in Eq.~\eqref{cqtchil} with
  $\lambda=h$, we find that, in the large-$L$ limit the fidelity
susceptibility diverges as
\begin{equation}
\chi_F(h,L)\sim L^{15/4}{\cal F}_2(\kappa), \qquad \kappa = h \, L^{15/8}.
\label{chifhl}
\end{equation}

On the other hand, in the usual setting considered in the
literature, the transverse field $g$ is varied across the value
$g=1$, and the longitudinal field is kept fixed at
$h=0$~\cite{ZP-06, CGZ-07, RD-11, Damski-13, LZW-18, CWHW-08}.
In such case, an analogous FSS follows~\cite{Damski-13},
where the scaling variable of Eq.~\eqref{lappaCQT} corresponding
to the transverse field is $\kappa_g = (g-1) L^{y_g}$. For the
quantum Ising chain, the RG dimension
\begin{equation}
  y_g=1/\nu,
\end{equation}
where $\nu=1$ (see again Ref.~\cite{CPV-14}). Therefore,
Eq.~\eqref{cqtchil} readily implies
\begin{equation}
  \chi_F(g,L)\sim L^2 {\cal F}_2^{(g)}(\kappa_g), \qquad \kappa_g = (g-1) L.
\end{equation}

\subsection{FSS at the first-order transition}
\label{sec:FOQT}

The FOQTs, occurring at $g<1$ along the line $h=0$, can be related to
the level crossing of the two lowest magnetized states $|+\rangle$ and
$|-\rangle$ for $h=0$, such that $\langle \pm | \sigma_i^{(3)}
| \pm \rangle = \pm \,m_0$, with $m_0 = (1 -
g^2)^{1/8}$~\cite{Pfeuty-70}.

Contrary to CQTs, the distinctive feature of FOQTs is a remarkable
qualitative dependence of their features on the boundary conditions.
As we shall see below in a variety of different situations in the
context of the Ising model, this also emerges in the FSS of the
fidelity susceptibility, exhibiting completely different scalings,
according to the size dependence of the energy difference of the
lowest energy levels.

\subsubsection{Periodic/open boundary conditions}
\label{sec:PBC-OBC}

\begin{figure}[!t]
  \includegraphics[width=0.95\columnwidth]{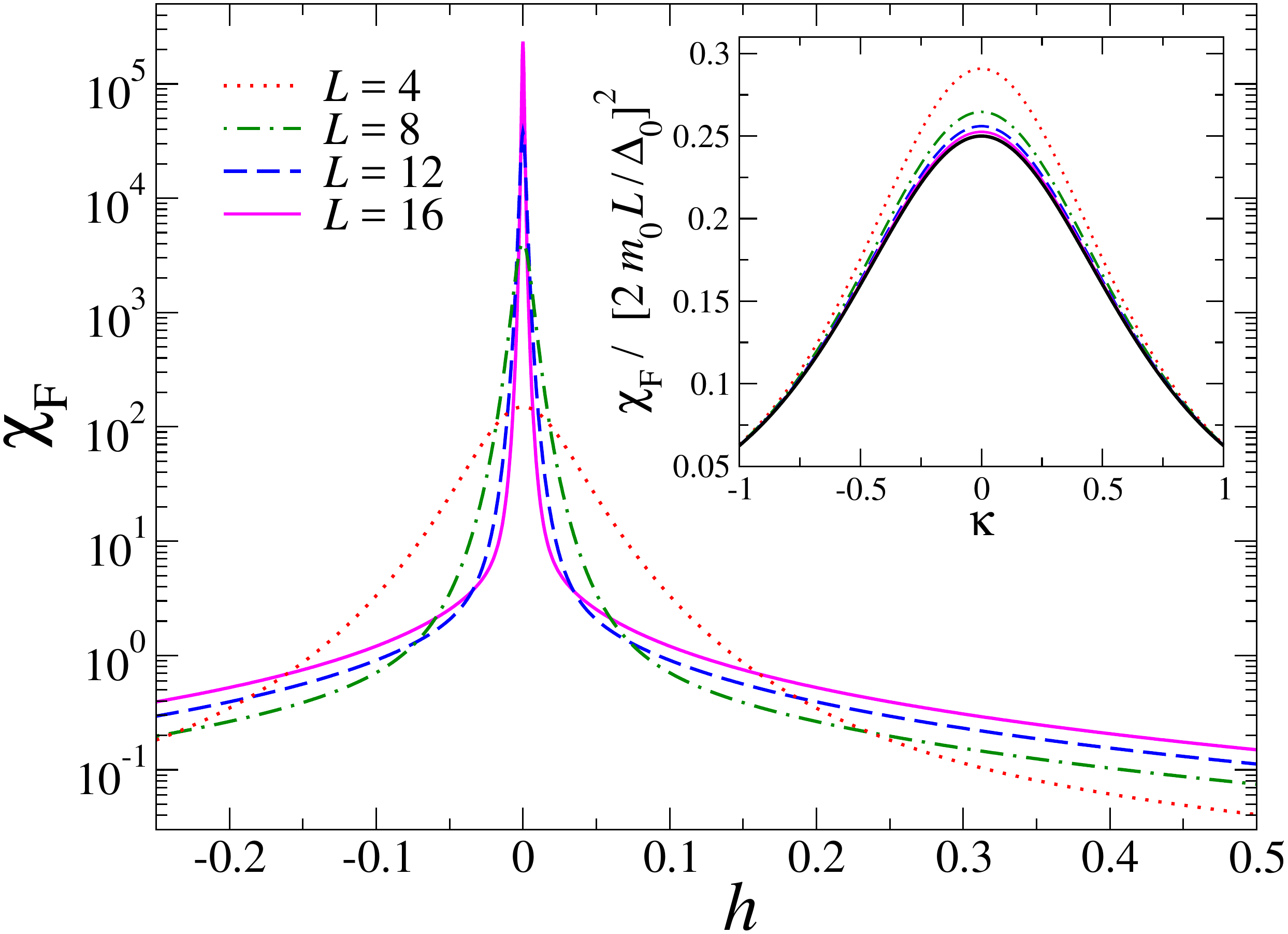}
  \caption{Fidelity susceptibility $\chi_F(h,L)$ for the Ising
    model~\eqref{hedef} with $g=0.9$ and PBC, associated with changes
    of the longitudinal parameter $h$, for some values of $L$, up to
    $L=16$.  The inset displays curves for $\chi_F/[2 m_0
      L/\Delta_0(L)]^2$, as a function of $\kappa=2 m_0 h L /
    \Delta_0(L)$ [see Eq.~\eqref{chiffo}], converging to the scaling
    function ${\cal F}_2^{(2l)}(\kappa)$ (thick black line),
    cf. Eq.~(\ref{f2l}). Analogous results are obtained for other
    value of $g<1$.}
  \label{FidSusc_PBC_g09}
\end{figure}

In a finite system of size $L$ with periodic or open boundary
conditions (PBC and OBC, respectively), the lowest states are
superpositions of $|+\rangle$ and $|-\rangle$, due to tunneling
effects. Their energy difference $\Delta_0(L) \sim g^L$ vanishes
exponentially with $L$.  More precisely~\cite{GJ-87}:
\begin{eqnarray}
  \Delta_0(L) = & 2 \, (1-g^2) \, g^L \, \big[ 1+ O(g^{2L}) \big] &
  \mbox{for OBC}, \\ \Delta_0(L) \approx & 2 \, \sqrt{(1-g^2) / (\pi
    L)} \, g^L \hspace*{0.85cm} & \mbox{for PBC}.
\end{eqnarray}
Conversely, the difference $\Delta_{0,i}\equiv E_i-E_0$ for higher
excited states ($i > 1$) remains finite for $L\to \infty$.  The
interplay of the size $L$ and the field $h$ gives rise to the FSS of
the low-energy properties~\cite{CNPV-14}. Its scaling variable is
obtained from Eq.~(\ref{kappafoqt}), i.e.
\begin{equation}
  \kappa = \frac{2 m_0 h L}{\Delta_0(L)},
  \label{kappafo}
\end{equation}
using the fact that $E_h=2m_0hL$ is the energy variation associated
with $h$.  The FSS limit corresponds to $L\to \infty$ and $h\to 0$,
keeping $\kappa$ fixed.  Correspondingly, the energy difference of the
lowest states and the magnetization behave as~\cite{CNPV-14}
$\Delta(h,L) \approx \Delta_0(L) \, {\cal D}(\kappa)$ and $m(L,h)
\approx m_0 \, {\cal M}(\kappa)$, where ${\cal D}(\kappa)$ and ${\cal
  M}(\kappa)$ are scaling functions independent of $g$.

The FSS of the fidelity and its susceptibility is given by
Eqs.~\eqref{fisca} and~\eqref{chifscal}. We obtain
\begin{equation}
  \chi_F(h,L) \approx \left[ {2 m_0 L\over \Delta_0(L)}\right]^2 
\!\!{\cal F}_2(\kappa),
\label{chiffo}
\end{equation} 
implying that it exponentially diverges with $L$. This is confirmed by
the numerical results~\cite{footnotenr} of Fig.~\ref{FidSusc_PBC_g09},
where the curves of $\chi_F$ for PBC display sharp, and exponentially
increasing, peaks around $h=0$, while $\chi_F=O(L)$ for larger $|h|$.

\begin{figure}[!t]
  \includegraphics[width=0.8\columnwidth]{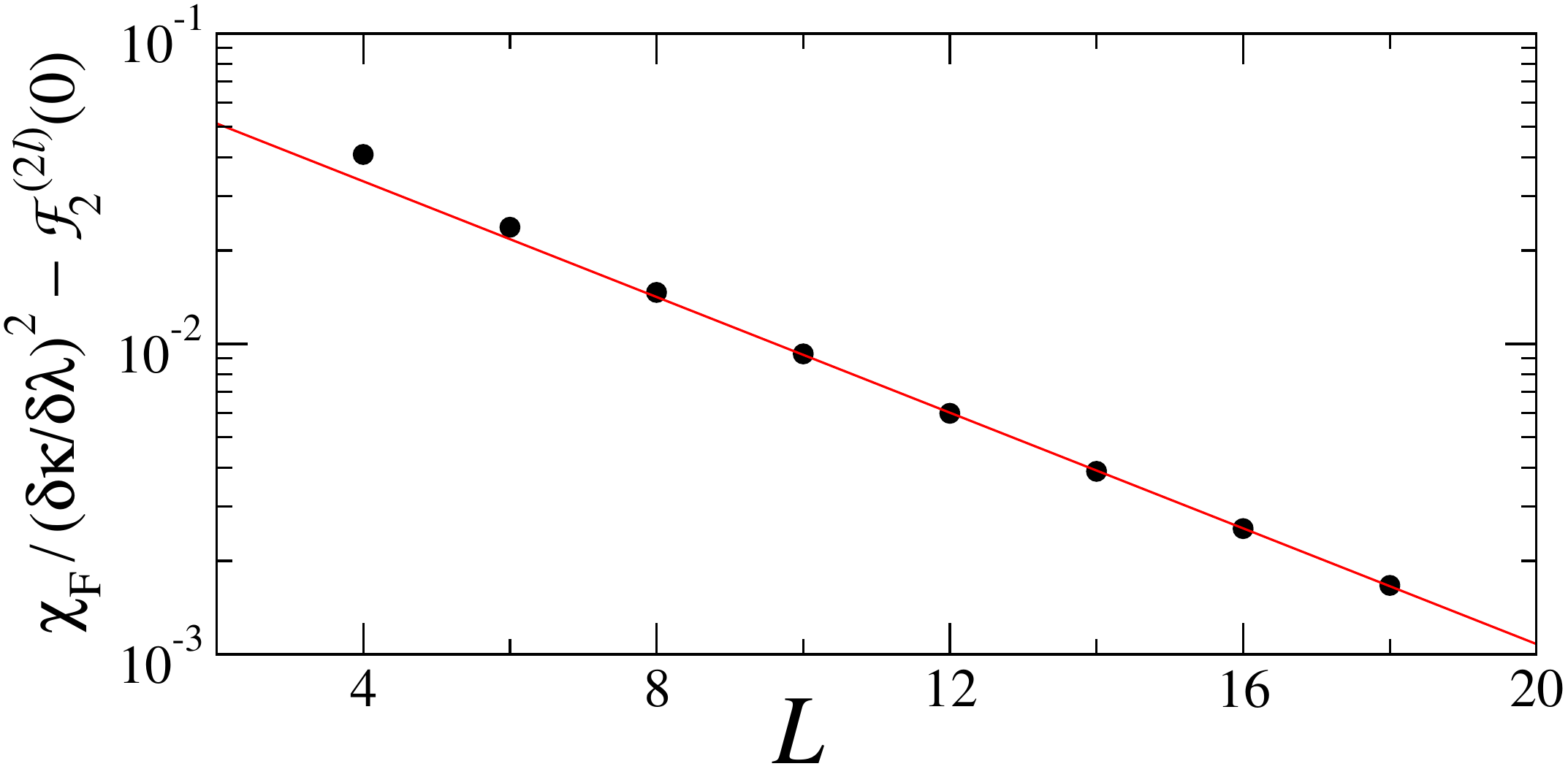}
  \caption{Convergence of the finite-size fidelity susceptibility to
    the asymptotic scaling function ${\cal F}_2(\kappa)$.  Data are
    for PBC, with $g=0.9$ and $\kappa=0$ (see
    Fig.~\ref{FidSusc_PBC_g09}).  We plot the rescaled susceptibility
    $\chi_F / [2 m_0 L / \Delta_0(L)]^2$ as a function of $L$,
    subtracting the asymptotic value given by ${\cal F}_2^{(2l)}(0) =
    1/4$.  The red line is an exponential fit of data for $5 \leq L
    \leq 18$.}
    \label{FiniteL_PBC}
\end{figure}

Since the low-energy spectrum for PBC and OBC across the FOQT is
characterized by the level crossing of the two lowest states, while
the energy differences with the other ones remain $O(1)$, the
asymptotic FSS can be exactly obtained by performing a two-level
truncation of the spectrum~\cite{CNPV-14, PRV-18, PRV-18b}, keeping
only the lowest energy levels $|\pm \rangle$.  Details are provided in
App.~\ref{app:2lev}, where an extension to finite temperature is also
presented, thus confirming Eq.~(\ref{fiscaT}).  The net result is
that, using the corresponding two-level effective Hamiltonian, we get
\begin{equation}
  {\cal F}^{(2l)}(\kappa,\delta\kappa) = {\rm cos}(\delta\alpha/2),
\end{equation}
where we defined
\begin{equation}
  \delta \alpha = \arctan{\Big[ \frac{1}{\kappa + \delta \kappa} \Big] }
  -\arctan{\frac{1}{\kappa}}
\end{equation}
with $\arctan[x] \in (0, \pi)$. Moreover
\begin{equation}
{\cal F}_{2}^{(2l)}(\kappa) = {1\over 4 (1 + \kappa^2)^2} .
\label{f2l}
\end{equation}
The inset of Fig.~\ref{FidSusc_PBC_g09} evidences the convergence of
$\chi_F/[2 m_0 L/\Delta_0(L)]^2$ to the scaling function ${\cal
  F}_2^{(2l)}(\kappa)$, as a function of the scaling variable $\kappa$
in Eq.~\eqref{kappafo}, which clearly turns out to be exponential, as
shown by Fig.~\ref{FiniteL_PBC}.

\begin{figure}[!t]
  \includegraphics[width=0.95\columnwidth]{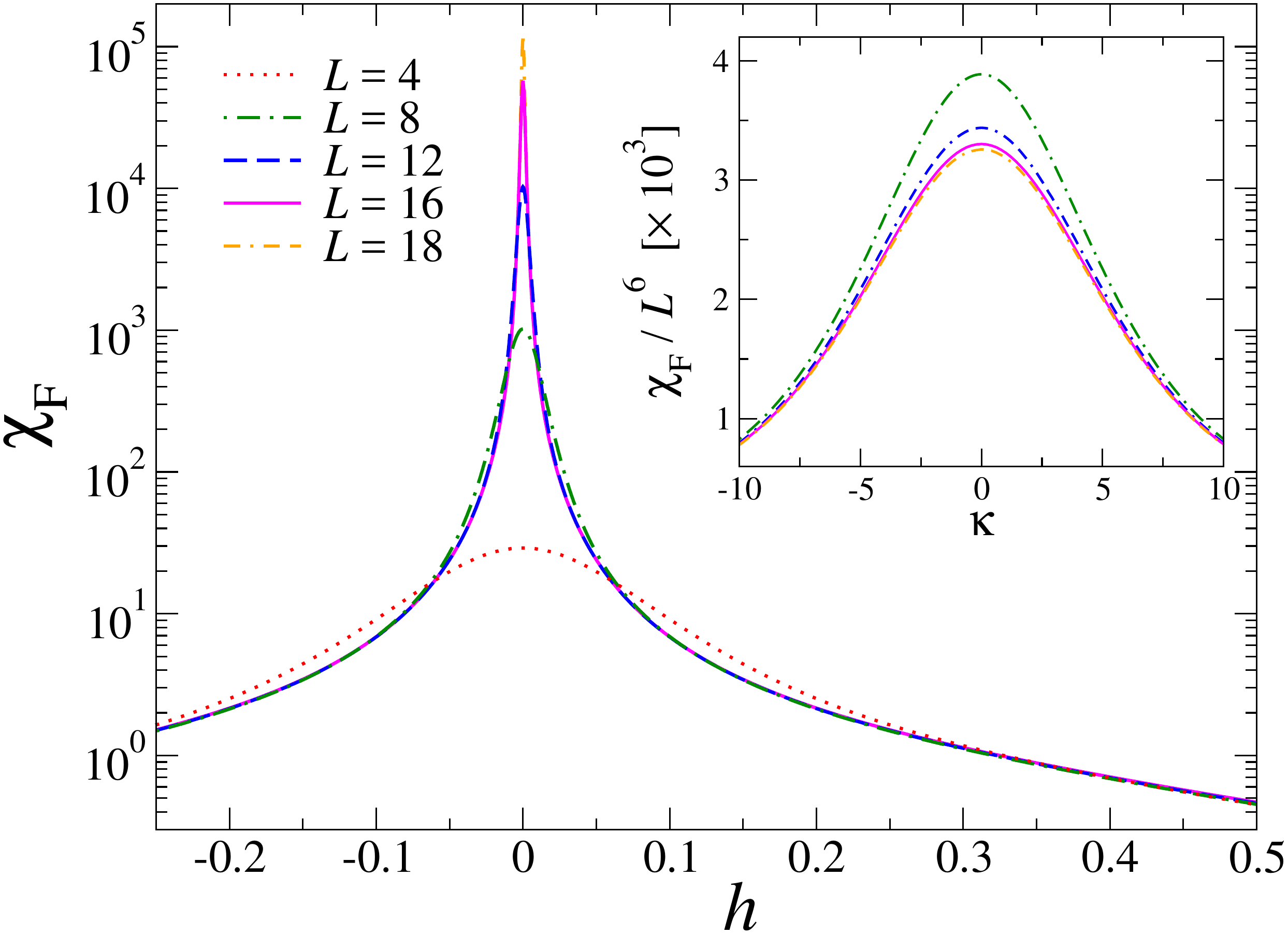}
  \caption{Same as in Fig.~\ref{FidSusc_PBC_g09}, but for $g=0.5$ and
    ABC.  The inset shows the rescaled fidelity susceptibility
    according to Eq.~\eqref{chifscalcqtabc}, for $\kappa=h L^3$.  The
    curves for $\chi_F/L^6$ clearly approach a scaling function of
    $\kappa$.}
\label{FidSusc_ABC_g05}
\end{figure}

\subsubsection{Antiperiodic boundary conditions}
\label{sec:ABC}

As already mentioned, the FOQT scenario based on the avoided crossing
of two levels, holding for PBC and OBC, is not always realized. 
Indeed a quite different behavior emerges when
considering antiperiodic boundary conditions (ABC). This is
essentially related to the fact that the corresponding low-energy
states are one-kink (nearest-neighbor pair of antiparallel spins)
states, behaving as one-particle states with $O(L^{-1})$
momenta. Thus, the energy difference of the lowest levels displays a
power-law behavior~\cite{GJ-87}:
\begin{equation}
\Delta_0(L) = [g/(1-g)] \, \pi^2\, L^{-2} + O(L^{-4}).
\label{deltalabc}
\end{equation}
Then, following Eq.~(\ref{kappafoqt}), we can define the scaling variable
\begin{equation}
\kappa = hL^3. 
\label{habc}
\end{equation}
Indeed, since the energy associated with the longitudinal field
$h$ scales as $E_h(h,L)\sim hL$, and the gap $\Delta_0(L)\sim
L^{-2}$, it is immediate to see that the ratio~(\ref{kappafoqt})
obeys the same dependence on $h$ and $L$ as in Eq.~\eqref{habc}.

\begin{figure}[!t]
  \includegraphics[width=0.8\columnwidth]{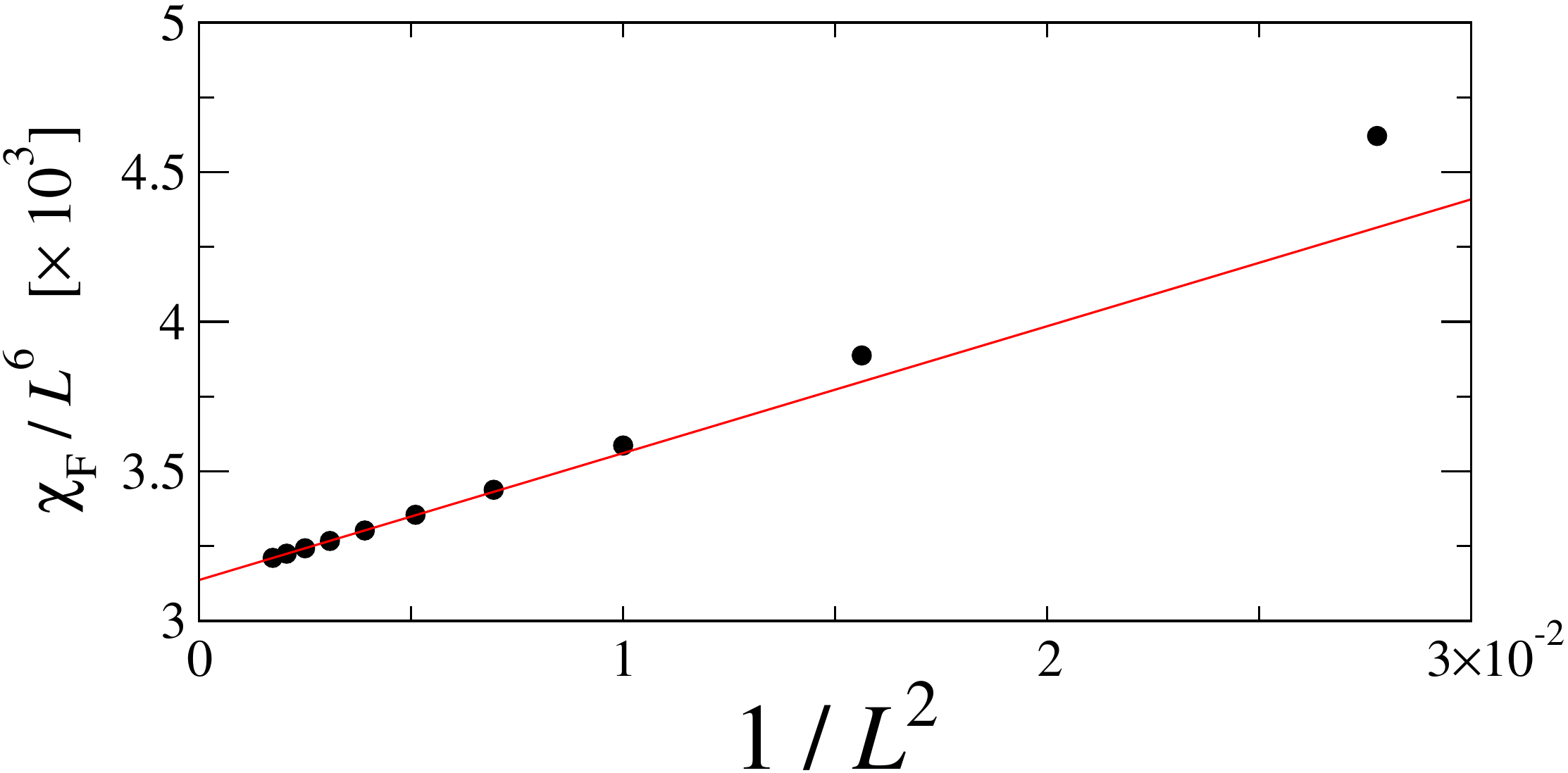}
  \caption{Same as in Fig.~\ref{FiniteL_PBC}, but for ABC, with
    $g=0.5$ and $\kappa=0$ (see Fig.~\ref{FidSusc_ABC_g05}).  Here we
    plot the rescaled susceptibility $\chi_F / L^6$ as a function of
    $L^{-2}$.  The red line is a power-law fit of data for $14 \leq L
    \leq 24$.}
    \label{FiniteL_ABC}
\end{figure}

The general ansatz (\ref{chiflal}) predicts a power-law behavior for
the fidelity susceptibility,
\begin{equation}
  \chi_F(h,L) \approx L^{6} \; {\cal F}_2^{(a)}(\kappa),
  \label{chifscalcqtabc}
\end{equation}
since $\partial \kappa/\partial h= L^3$.  Again, this FSS is nicely
supported by the numerical data~\cite{footnotenr} of
Fig.~\ref{FidSusc_ABC_g05}.  With increasing $L$, the curves for the
ratio $\chi_F/L^6$ appear to approach a scaling function ${\cal
  F}_2^{(a)}(\kappa)$.  Finite-size corrections appear to be
power-law, of the order $O(L^{-2})$, as is visible from
Fig.~\ref{FiniteL_ABC}.

It is important to emphasize that, unlike the cases of PBC and OBC,
for ABC the scaling functions cannot be obtained by a two-level
approximation, because the low-energy spectrum at the transition point
presents a tower of excited stated with $\Delta_{0,i}=O(L^{-2})$.  We
also note that, for $|h|>0$, $\chi_F$ appears to converge to a finite
value with increasing $L$, unlike for PBC (compare the tails of
Fig.~\ref{FidSusc_ABC_g05} with those of Fig.~\ref{FiniteL_PBC}).

\subsubsection{Equal and fixed boundary conditions}
\label{sec:EFBC}

Let us finally consider equal fixed boundary conditions (EFBC)
favoring one of the two magnetized phases.  This is obtained by adding
equal fixed spin states $|\!\downarrow\rangle$ at the ends $x=0$ and
$x=L+1$ of the chain (\ref{hedef}).  In such case, the interplay
between the size $L$ and the bulk field $h$ gives rise to a more
complex finite-size behavior with respect to that of neutral boundary
conditions, such as PBC and ABC~\cite{PRV-18c}.

When $h=0$, the system is in the negatively magnetized phase, and
$\Delta_0(L) = 4 (1-g) + O(L^{-2})$.  For sufficiently small $h$, the
observables depend smoothly on it. Then the system undergoes a sharp
transition to the other phase at $h \approx h_{tr}(L)>0$, which tends
to zero with increasing $L$, asymptotically as $h_{tr}(L)\approx
\eta(g)/L$, where $\eta(g)$ is a $g$-dependent
constant~\cite{PRV-18c}.  This sharp transition corresponds to the
minimum $\Delta_m(L)$ of the energy difference $\Delta(h,L)$ of the
lowest levels, which vanishes exponentially with increasing $L$, as
$\Delta_m(L)\sim e^{-b(g)L}$.  Around $h_{tr}$, the suitable scaling
variable turns out to be
\begin{equation}
\kappa = {[h-h_{tr}(L)] L\over \Delta_m(L)},
\label{kappafbc}
\end{equation}
analogously to that of PBC and OBC, apart from the $1/L$ shift of the
transition point.  The corresponding scaling behaviors, $\Delta(h,L)
\approx \Delta_m(L) \, {\cal D}(\kappa)$ and $m(h,L) \approx m_0 \,
        {\cal M}(\kappa)$, turn out to be those emerging from an
        avoided two-level crossing, similarly to the case of PBC.

\begin{figure}[!t]
  \includegraphics[width=0.95\columnwidth]{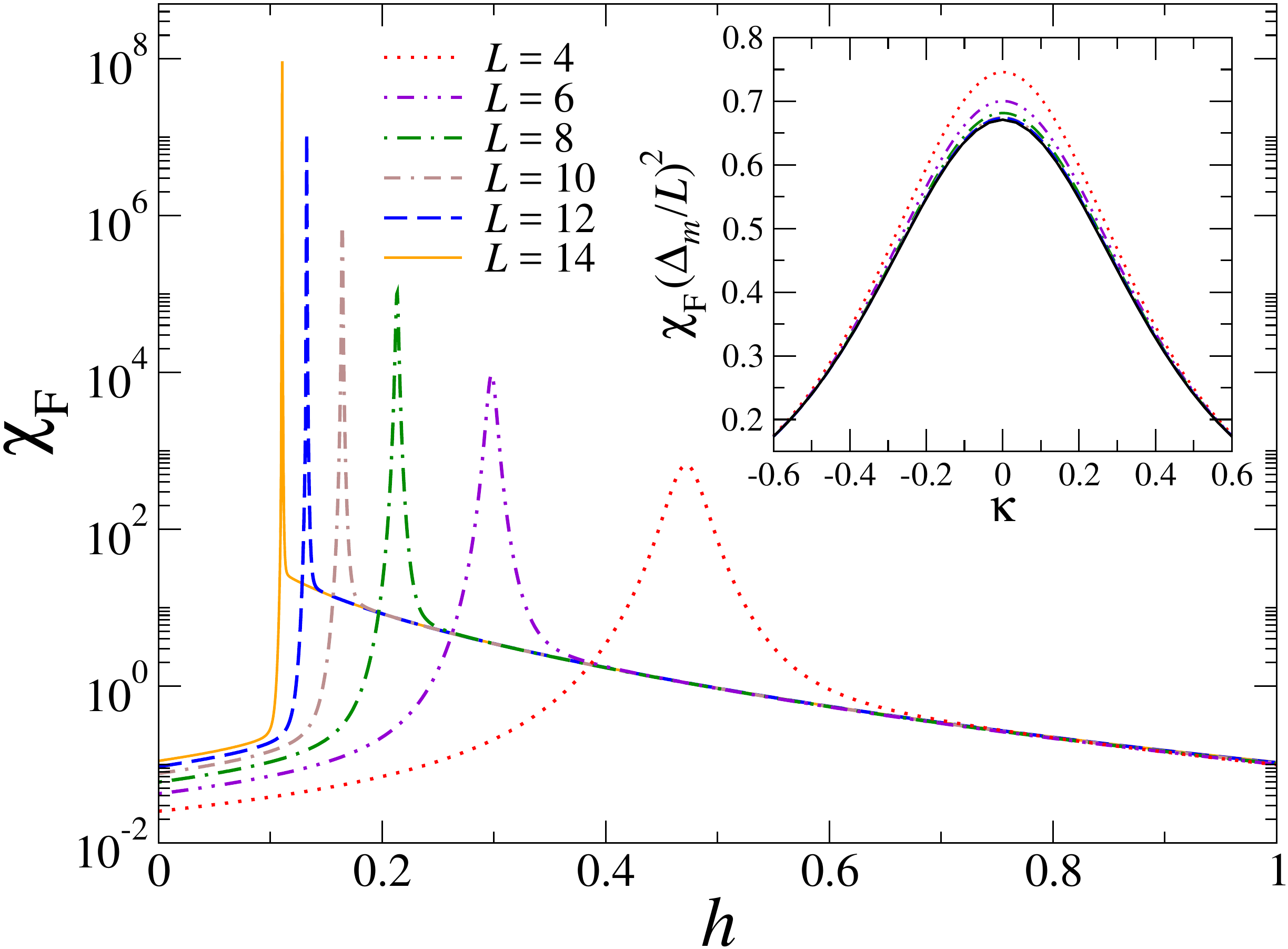}
  \caption{Same as in Fig.~\ref{FidSusc_PBC_g09}, but for $g=0.5$ and
    EFBC.  The inset shows the FSS of $\chi_F$ around $h=h_{tr}(L)$,
    with $\kappa = [h-h_{tr}(L)] L / \Delta_m(L)$.  With increasing
    $L$, the curves of $(\Delta_m/L)^{2} \chi_F(h,L)$ rapidly
    (exponentially) approach the two-level scaling function, cf.
    Eq.~\eqref{chifscalcqtefbc}, with $a \approx 0.67$, $b \approx
    1.64$.}
    \label{FidSusc_EFBC_g05}
\end{figure}

Figure~\ref{FidSusc_EFBC_g05} shows the $h$-dependence of the fidelity
susceptibility $\chi_F(h,L)$ for several values of
$L$~\cite{footnotenr}.  Its behavior reflects that of other
observables.  In particular it is smooth around $h=0$, since we
checked that the ratio $\chi_F(h,L)/\chi_F(h=0,L)$ rapidly approaches
a function of $h$ only, with $\chi_F(0,L)= O(L)$ (not shown).  Then,
with increasing $h$, the curves show a sharp peak around $h_{tr}(L)$,
whose maximum rapidly increases with $L$, and becomes narrower and
narrower.  For even larger $h$, $\chi_F(h,L)$ tends to rapidly become
independent of $L$; this is related to the fact that the ground state
is essentially given by spatially separated kink and antikink
structures, whose position depends smoothly on $h$~\cite{PRV-18c}.
The scaling behavior around $h_{tr}(L)$ can be inferred from the
general ansatz~(\ref{chifscal}):
\begin{equation}
{\Delta_m^2\over L^2} \,\chi_F(h,L) \approx a \, {\cal
  F}_2^{(2l)}(b\,\kappa),
\label{chifscalcqtefbc}
\end{equation}
where the scaling variable $\kappa$ is that given in
Eq.~(\ref{kappafbc}), and ${\cal F}_2^{(2l)}(x)$ is the two-level
scaling function (\ref{f2l}), while $a$ and $b$ are appropriate
normalizations.  This is confirmed by numerical data in the inset of
Fig.~\ref{FidSusc_EFBC_g05}.

Finally we mention that the case of fixed, but opposite, boundary
conditions (OFBC), i.e. $|\!\downarrow\rangle$ and
$|\!\uparrow\rangle$ at the ends of the chain, is supposed to be
similar to that with ABC~\cite{CPV-15, CPV-15b}, because the
low-energy states are again one-kink states.  Thus, $\Delta_0(L)\sim
L^{-2}$ as well, and a power-law behavior such as
(\ref{chifscalcqtabc}) is expected.

\section{Summary and conclusion}
\label{sec:concl}

In conclusion, we have shown that the ground-state fidelity and the
corresponding susceptibility develop FSS behaviors at both CQTs and
FOQTs, arising from the interplay between the driving parameter and
the system size.  At CQTs the fidelity susceptibility generally shows
power laws: $\chi_F \sim L^{2y_\lambda}$ at the transition point,
where $y_\lambda$ is the universal exponent associated with the
critical properties of the corresponding Hamiltonian perturbation.  At
CQTs boundary conditions only affect the scaling functions of
observables.  This sharply contrasts with FOQTs, whose distinctive
feature is a remarkable qualitative dependence on the boundary
conditions; indeed the fidelity susceptibility may show exponential or
power-law FSS, essentially related to the size dependence of the
energy difference of the lowest levels.  In particular, exponential
behaviors develop for boundary conditions such as PBC or EFBC, for
which $\chi_F \sim e^{c L^d}$ at the transition point. Conversely,
power-law behaviors similar to those occurring at CQTs turn out to
develop for ABC or for OFBC.
Our findings have been confirmed by analytical and numerical results
for the one-dimensional quantum Ising model.

It is worth mentioning that the FSS treatment adopted here
for the study of the ground-state fidelity in the quantum Ising ring
shares important similarities with the approach previously employed
to address other quantities in different kinds of QTs.
First of all, the definition of the relevant scaling variable $\kappa$ 
through Eq.~\eqref{lappaCQT} and Eq.~\eqref{kappafoqt}
(for first-order and for continuous QTs, respectively),
is closely related to the general arguments put forward
in Refs.~\cite{SGCS-97, CPV-14} for CQTs, and in Ref.~\cite{CNPV-14} for FOQTs.
Moreover, the striking dependence of the FSS behavior at FOQTs (here evidenced for
the fidelity) has been spotlighted in similar contexts as well, for low-lying energy gaps,
local observables, and correlation functions~\cite{CPV-15, CPV-15b, PRV-18c},
yielding consistent results.

All these connections are in support of the broad validity of our FSS theory:
indeed we expect it to hold even in higher
dimensions and for FOQTs of other models, where it would be tempting
to have a direct numerical validation.  Moreover, the possibility to
generalize it to finite temperature makes it relevant also to quantum
thermometry close to criticality, where estimation performances depend
on the scaling behavior~\cite{ZPC-08}.
We also notice that the FSS frameworks have been extended to the off-equilibrium quantum dynamics,
focusing on both time-dependent perturbations~\cite{PRV-18} and sudden
quenches~\cite{PRV-18b}. 
By defining scaling variables that are consistent with the procedure considered in this paper,
and including further ones associated with the time and the dynamic variables,
dynamic FSS behaviors have been shown to emerge even in other contexts,
as for the decoherence properties~\cite{Vicari-18} and the statistics of the work~\cite{NRV-18}.

As suggested from the present study, the FSS of the fidelity is
amenable to a direct experimental verification by means of small-size
quantum simulators (i.e., of the order of ten spins), which can thus
serve as a probe of the nature of the transition itself.  A possible
strategy would be to measure the Loschmidt echo after a sudden
quench~\cite{ZPRS-08, ZCCL-09}, a quantity strictly related to the
fidelity susceptibility~\cite{GPSZ-06, JP-09, DGP-10, SA-17}, which
might shed light on the mutual interplay between QTs, entanglement and
decoherence~\cite{QSZS-06, RCGMF-07, HGMPM-12}.

\acknowledgments
We thank R. Fazio and A. Pelissetto for fruitful discussions.

\appendix

\section{Two-level reduction of the spectrum across the FOQT line}
\label{app:2lev}

As stated in Sec.~\ref{sec:PBC-OBC}, in the thermodynamic limit, the
low-energy spectrum for PBC and for OBC across the FOQT is
characterized by the level crossing of the two lowest states, while
the energy differences with the other ones remain finite.  The
asymptotic FSS behavior for the fidelity and for its susceptibility
can be thus exactly obtained by performing a two-level truncation of
the spectrum, following Refs.~\cite{CPV-14, PRV-18, PRV-18b}, keeping
only the lowest energy levels.  For the sake of completeness, here we
sketch this derivation.

The effective Hamiltonian, written in the Hilbert space spanned by the
two lowest magnetized states $|+ \rangle$ and $|-\rangle$ for $h=0$,
i.e.~such that $\langle \pm | \sigma_i^{(3)} | \pm \rangle =
\pm \,m_0$ [with $m_0 = (1 - g^2)^{1/8}$], reads:
\begin{equation}
  H_{2}(h) = - \beta \, \sigma^{(3)} + \delta \, \sigma^{(1)} .
\label{hrtds}
\end{equation}
The parameters $\beta$ and $\delta$ correspond to $\beta = m_0 h L$ and
$\delta={\Delta_0/2}$, such that $\kappa(h)= \beta/\delta$.  The
eigenstates are
\begin{eqnarray}
&&|0\rangle = \sin(\alpha/2) \, |-\rangle -  
\cos(\alpha/2) \, |+\rangle, \label{eigstate0la}\\
&& |1\rangle =  \cos(\alpha/2) |-\rangle +
\sin(\alpha/2) \, |+\rangle, \label{eigstate1la}
\end{eqnarray}
where $\tan \alpha = \kappa^{-1}$ with $\alpha \in (0,\pi)$, and $E_1
- E_0 = \Delta_0 \; \sqrt{1 + \kappa^2}$.  

Straightforward calculations confirm the FSS behavior in Eq.~\eqref{fisca}
of the zero-temperature fidelity:
$F(\lambda,\delta \lambda, L) \approx {\cal F}(\kappa, \delta \kappa)$.
Indeed, we obtain
\begin{equation}
  F(h,\delta h,L) \approx {\cal F}^{(2l)}(\kappa,\delta\kappa) = {\rm
    cos}(\delta\alpha/2),
\label{fpbcsca}
\end{equation}
where $\tan(\alpha + \delta\alpha) = (\kappa + \delta\kappa)^{-1}$.
The corresponding scaling function in Eq.~\eqref{f2l} of the fidelity
susceptibility is thus easily obtained.

\begin{figure}[!t]
  \includegraphics[width=0.95\columnwidth]{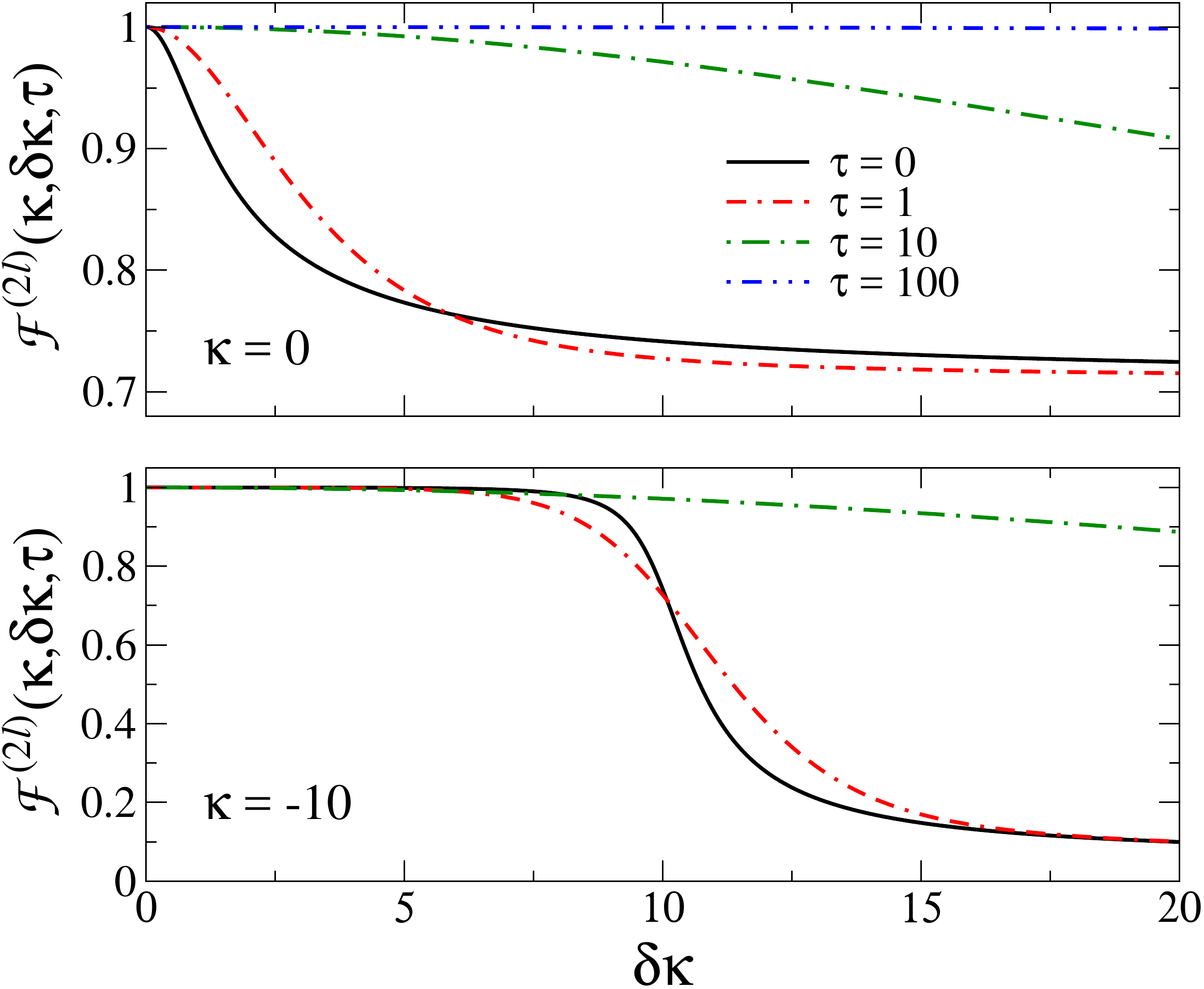}
  \caption{Scaling function of the fidelity susceptibility
    for two different values of $\kappa$, at finite temperature $\tau$,
    as obtained in a two-level truncation scheme.
    The continuous black curves correspond to the zero-temperature case,
    for which the analytic curve of Eq.~\eqref{fpbcsca} holds.}
  \label{fidelity2l}
\end{figure}

As discussed in Sec.~\ref{sec:FSS_fid}, the definition of fidelity can
be extended to finite temperature as well, through
Eq.~\eqref{fiddefT}.  The computation based on the two-level
truncation confirms the FSS behavior put forward in
Eq.~\eqref{fiscaT}.  In Fig.~\ref{fidelity2l} we report some plots of
the scaling function ${\cal F}^{(2l)}(\kappa,\delta\kappa,\tau)$, for
different values of $\kappa$ and $\tau$.  Note that, for $\kappa=0$,
the zero-temperature fidelity at large $\delta \kappa$ approaches the
asymptotic value $|\langle +|0\rangle| = 2^{-1/2} \approx 0.707$.  On
the other hand, for $\kappa \to -\infty$, it approaches zero, since it
corresponds to abruptly sweeping from one side of the transition, to
the other one.  The effect of the temperature is to progressively
smoothen the behavior of the various curves with $\delta \kappa$.


\newpage

\begin{thebibliography}{99}

\bibitem{Sachdev-book}
  S. Sachdev, {\em Quantum Phase Transitions},
  (Cambridge University, Cambridge, England, 1999).

\bibitem{SGCS-97} S. L. Sondhi, S. M. Girvin, J. P. Carini, and
  D. Shahar, Continuous quantum phase transitions,
  Rev. Mod. Phys. {\bf 69}, 315 (1997).

\bibitem{AFOV-08}
  L. Amico, R. Fazio, A. Osterloh, and V. Vedral,
  Entanglement in many-body systems,
  Rev. Mod. Phys. {\bf 80}, 517 (2008).

\bibitem{Gu-10}
  S.-J. Gu, Fidelity approach to quantum phase transitions,
  Int. J. Mod. Phys. B {\bf 24}, 437 (2010).

\bibitem{BAB-17} D. Braun, G. Adesso, F. Benatti, R. Floreanini,
  U. Marzolino, M. W. Mitchell, and S. Pirandola, Quantum-enhanced
  measurements without entanglement, 
  Rev. Mod. Phys. {\bf 90}, 035006 (2018).

\bibitem{Anderson-67} P. W. Anderson, Infrared catastrophe in Fermi
  gases with local scattering potentials, Phys. Rev. Lett. {\bf 18},
  1049 (1967).

\bibitem{BC-94} S. L. Braunstein and C. M. Caves, Statistical distance
  and the geometry of quantum states, Phys. Rev. Lett. {\bf 72}, 3439
  (1994).
  
\bibitem{Paris-09} M. G. A. Paris, Quantum estimation for quantum
  technology, Int. J. Quant. Inf. {\bf 7}, 125 (2009).

\bibitem{ZPC-08} P. Zanardi, M. G. A. Paris, and L. Campos Venuti,
  Quantum criticality as a resource for quantum estimation,
  Phys. Rev. A {\bf 78}, 042105 (2008).

\bibitem{IKCP-08} C. Invernizzi, M. Korbman, L. Campos Venuti, and
  M. G. A. Paris, Optimal quantum estimation in spin systems at
  criticality, Phys. Rev. A {\bf 78}, 042106 (2008).
  
\bibitem{ZP-06}
  P. Zanardi and N. Paunkovi\'c,
  Ground state overlap and quantum phase transitions,
  Phys. Rev. E {\bf 74}, 031123 (2006).

\bibitem{YLG-07} W. L. You, Y. W. Li, and S. J. Gu, Fidelity, dynamic
  structure factor, and susceptibility in critical phenomena,
  Phys. Rev. E {\bf 76}, 022101 (2007).

\bibitem{VZ-07} L. Campos Venuti and P. Zanardi, Quantum Critical
  Scaling of the Geometric Tensors, Phys. Rev. Lett. {\bf 99}, 095701
  (2007).
  
\bibitem{CGZ-07} M. Cozzini, P. Giorda, and P. Zanardi, Quantum phase
  transitions and quantum fidelity in free fermion graphs,
  Phys. Rev. B {\bf 75}, 014439 (2007).

\bibitem{RD-11} M. M. Rams and B. Damski, Quantum fidelity in the
  thermodynamic limit, Phys. Rev. Lett. {\bf 106}, 055701 (2011);
  Scaling of ground state fidelity in the thermodynamic limit: XY
  model and beyond, Phys. Rev. A {\bf 84}, 032324 (2011).

\bibitem{MPD-11} V. Mukherjee, A. Polkovnikov, and A. Dutta,
  Oscillating fidelity susceptibility near a quantum multicritical
  point, Phys. Rev. B {\bf 83}, 075118 (2011).
  
\bibitem{Damski-13} B. Damski, Fidelity susceptibility of the quantum
  Ising model in the transverse field: The exact solution,
  Phys. Rev. E {\bf 87}, 052131 (2013).

\bibitem{LZW-18} Q. Luo, J. Zhao, and X. Wang, Fidelity susceptibility
  of the anisotropic $XY$ model: The exact solution,
  Phys. Rev. E {\bf 98}, 022106 (2018).
  
\bibitem{CWHW-08} S. Chen, L. Wang, Y. Hao, and Y. Wang, Intrinsic
  relation between ground-state fidelity and the characterization of a
  quantum phase transition, Phys. Rev. A {\bf 77}, 032111 (2008).
  
\bibitem{SAC-09} D. Schwandt, F. Alet, and S. Capponi, Quantum Monte
  Carlo simulations of fidelity at magnetic quantum phase transitions,
  Phys. Rev. Lett. {\bf 103}, 170501 (2009).

\bibitem{LLZ-09} B. Li, S.-H. Li, and H.-Q. Zhou, Quantum phase
  transitions in a two-dimensional quantum XYX model: Ground-state
  fidelity and entanglement, Phys. Rev. E {\bf 79}, 060101(R) (2009).

\bibitem{AASC-10} A.F. Albuquerque, F. Alet, C. Sire, and S. Capponi,
  Quantum Critical Scaling of Fidelity Susceptibility, Phys. Rev. B
  {\bf 81}, 064418 (2010).

\bibitem{Sirker-10} J. Sirker, Finite-Temperature Fidelity
  Susceptibility for One-Dimensional Quantum Systems,
  Phys. Rev. Lett. {\bf 105}, 117203 (2010).

\bibitem{Nishiyama-13} Y. Nishiyama, Criticalities of the transverse-
  and longitudinal-field fidelity susceptibilities for the $d = 2$
  quantum Ising model, Phys. Rev. E {\bf 88}, 012129 (2013).

\bibitem{SKV-14} G. Sun, A. K. Kolezhuk, and T. Vekua, Fidelity at
  Berezinskii-Kosterlitz-Thouless quantum phase transitions,
  Phys. Rev. B {\bf 91}, 014418 (2015).

\bibitem{BV-07} P. Buonsante and A. Vezzani, Ground-State Fidelity and
  Bipartite Entanglement in the Bose-Hubbard Model,
  Phys. Rev. Lett. {\bf 98}, 110601 (2007).

\bibitem{MHCFR-11} S. R. Manmana, K. R. A. Hazzard, G. Chen,
  A. E. Feiguin, and A. M. Rey, SU(N) magnetism in chains of ultracold
  alkaline-earth-metal atoms: Mott transitions and quantum
  correlations, Phys. Rev. A {\bf 84}, 043601 (2011).

\bibitem{CMR-13} J. Carrasquilla, S. R. Manmana, and M. Rigol, Scaling
  of the gap, fidelity susceptibility, and Bloch oscillations across
  the superfluid to Mott insulator transition in the one-dimensional
  Bose-Hubbard model, Phys. Rev. A {\bf 87}, 043606 (2013).

\bibitem{WLIMT-15} L. Wang, Y.-H. Liu, J. Imri\v ska, P. N. Ma, and
  M. Troyer, Fidelity susceptibility made simple: A unified quantum
  Monte Carlo approach, Phys. Rev. X {\bf 5}, 031007 (2015).

\bibitem{HWWW-16} L. Huang, Y. Wang, L. Wang, and P. Werner, Detecting
  phase transitions and crossovers in Hubbard models using the
  fidelity susceptibility, Phys. Rev. B {\bf 94}, 235110 (2016).

\bibitem{Kettemann-16} S. Kettemann, Exponential orthogonality
  catastrophe at the Anderson metal-insulator transition,
  Phys. Rev. Lett. {\bf 117}, 146602 (2016).

\bibitem{SS-17} S. Santhosh Kumar and S. Shankaranarayanan, Evidence
  of quantum phase transition in real-space vacuum entanglement of
  higher derivative scalar quantum field theories, Sci. Rep. {\bf 7},
  15774 (2017).

\bibitem{YGSL-08} S. Yang, S.-J. Gu, C.-P. Sun, and H.-Q. Lin,
  Fidelity susceptibility and long-range correlation in the Kitaev
  honeycomb model, Phys. Rev. A {\bf 78}, 012304 (2008).
  
\bibitem{OS-14} T. P. Oliveira and P. D. Sacramento, Entanglement
  modes and topological phase transitions in superconductors,
  Phys. Rev. B {\bf 89}, 094512 (2014).

\bibitem{KLS-16} E. J. K\"onig, A. Levchenko, and N. Sedlmayr,
  Universal fidelity near quantum and topological phase transitions in
  finite one-dimensional systems, Phys. Rev. B {\bf 93}, 235160
  (2016).

\bibitem{BGZ-14} L. Banchi, P. Giorda, and P. Zanardi, Quantum
  information-geometry of dissipative quantum phase transitions,
  Phys. Rev. E {\bf 89}, 022102 (2014).
  
\bibitem{MP-17} U. Marzolino and T. Prosen, Fisher information
  approach to nonequilibrium phase transitions in a quantum XXZ spin
  chain with boundary noise, Phys. Rev. B {\bf 96}, 104402 (2017).

\bibitem{PPBWJ-99} V. Piazza, V. Pellegrini, F. Beltram,
  W. Wegscheider, T. Jungwirth, and A. H. MacDonald, First-order phase
  transitions in a quantum Hall ferromagnet, Nature (London) {\bf
    402}, 638 (1999).
  
\bibitem{VBKN-99} T. Vojta, D. Belitz, T. R. Kirkpatrick, and
  R. Narayanan, Quantum critical behavior of itinerant ferromagnets,
  Ann. Phys. (Leipzig) {\bf 8}, 593 (1999).

\bibitem{UPH-04}
  M. Uhlarz, C. Pfleiderer, and S. M. Hayden,
  Quantum phase transitions in the itinerant ferromagnet ZrZn${}_2$,
  Phys. Rev. Lett. {\bf 93}, 256404 (2004).

\bibitem{Pfleiderer-05} C. Pfleiderer, Why first order quantum phase
  transitions are interesting, J. Phys. Condens. Matter {\bf 17}, S987
  (2005).

\bibitem{KRLF-09} W. Knafo, S. Raymond, P. Lejay, and J. Flouquet,
  Antiferromagnetic criticality at a heavy-fermion quantum phase
  transition, Nat. Phys. {\bf 5}, 753 (2009).

\bibitem{JKKM-08}
  T. J\"org, F. Krzakala, J. Kurchan, and A. C. Maggs,
  Simple glass models and their quantum annealing,
  Phys. Rev. Lett. {\bf 101}, 147204 (2008).
  
\bibitem{JLSZ-10} T. J\"org, F. Krzakala, G. Semerjian, and
  F. Zamponi, First-order transitions and the performance of quantum
  algorithms in random optimization problems, Phys. Rev. Lett. {\bf
    104}, 207206 (2010).

\bibitem{YKS-10} A. P. Young, S. Knysh, and V. N. Smelyanskiy, First
  order phase transition in the quantum adiabatic algorithm,
  Phys. Rev. Lett. {\bf 104}, 020502 (2010).

\bibitem{SN-12} Y. Seki and H. Nishimori, Quantum annealing with
  antiferromagnetic fluctuations, Phys. Rev. E {\bf 85}, 051112
  (2012).

\bibitem{CPV-14} M. Campostrini, A. Pelissetto, and E. Vicari,
  Finite-size scaling at quantum transitions, Phys. Rev. B {\bf 89},
  094516 (2014).

\bibitem{CNPV-14} M. Campostrini, J. Nespolo, A. Pelissetto, and
  E. Vicari, Finite-size scaling at first-order quantum transitions,
  Phys. Rev. Lett. {\bf 113}, 070402 (2014);
  Finite-size scaling at first-order quantum transitions of
  quantum Potts chains, Phys. Rev. E {\bf 91}, 052103 (2015).

\bibitem{PRV-18} A. Pelissetto, D. Rossini, and E. Vicari,
  Off-equilibrium dynamics driven by localized time-dependent
  perturbations at quantum phase transitions, Phys. Rev. B {\bf 97},
  094414 (2018).

\bibitem{PRV-18b} A. Pelissetto, D. Rossini, and E. Vicari, Dynamic
  finite-size scaling after a quench at quantum transitions,
  Phys. Rev. E {\bf 97}, 052148 (2018).

\bibitem{CPV-15} M. Campostrini, A. Pelissetto, and E. Vicari, Quantum
  transitions driven by one-bond defects in quantum Ising rings,
  Phys. Rev. E {\bf 91}, 042123 (2015).

\bibitem{CPV-15b}
  M. Campostrini, A. Pelissetto, and E. Vicari,
  Quantum Ising chains with boundary terms, J. Stat. Mech. (2015) P11015.

\bibitem{PRV-18c} A. Pelissetto, D. Rossini, and E. Vicari,
  Finite-size scaling at first-order quantum transitions when boundary
  conditions favor one of the two phases,
  Phys. Rev. E {\bf 98}, 032124 (2018).

\bibitem{CCD-09}
  Entanglement entropy in extended systems, edited by P. Calabrese,
  J. Cardy, and B. Doyon, J. Phys. A {\bf 42}, 500301 (2009).

\bibitem{DS-18}
  G. De Chiara and A. Sanpera, Genuine quantum correlations in
  quantum many-body systems: a review of recent progress,
  Rep. Prog. Phys. {\bf 81}, 074002 (2018).

\bibitem{Barber-83} M. N. Barber, Finite-size scaling, in {\em Phase
  transitions and critical phenomena}, vol. 8, page 145, C. Domb and
  J. L. Lebowitz eds.  (Academic Press, London 1983).

\bibitem{Privman-90} {\em Finite Size Scaling and Numerical
  Simulations of Statistical Systems}, ed. V. Privman (World
  Scientific, 1990).

\bibitem{PV-02} 
  A. Pelissetto and E. Vicari, 
  Critical phenomena and renormalization-group theory,
  Phys. Rep. {\bf 368}, 549 (2002).

\bibitem{GKMD-08} 
  F. M. Gasparini, M. O. Kimball, K. P. Mooney, and M. Diaz-Avilla,
  Finite-size scaling of He$^4$ at the superfluid transition,
  Rev. Mod. Phys. {\bf 80}, 1009 (2008).

\bibitem{BDZ-08} I. Bloch, J. Dalibard, and W. Zwerger, Many-body
  physics with ultracold gases, Rev.\ Mod.\ Phys.\ {\bf 80}, 885
  (2008).

\bibitem{FB-72} M. E. Fisher and M. N. Barber, Scaling Theory for
  Finite-Size Effects in the Critical Region, Phys. Rev. Lett. {\bf
    28}, 1516 (1972).

\bibitem{footnote1} To compare with Refs.~\cite{VZ-07, SAC-09}, simply
  use the scaling relation~\cite{Sachdev-book}: $y_\lambda = d + z -
  y_{H_I}$, where $z$ is the dynamic exponent, and $y_{H_I}$ is the RG
  dimension of $H_I/L^d$.

\bibitem{Uhlmann-76} A. Uhlmann, The ``transition probability'' in the
  state space of a ${}^\star$-algebra, Rep. Math. Phys. {\bf 9}, 273
  (1976).
  
\bibitem{Pfeuty-70} P. Pfeuty, The one-dimensional Ising model with a
  transverse field, Ann. Phys. {\bf 57}, 79 (1970).

\bibitem{GJ-87} G. G. Cabrera and R. Jullien, Role of boundary
  conditions in the finite-size Ising model, Phys. Rev. B {\bf 35},
  7062 (1987).

\bibitem{footnotenr} Numerical results have been obtained by means of
  exact diagonalization (for $L \leq 12$) or through a Lanczos
  algorithm (for $L \geq 14$).  The susceptibility has been extracted
  by quadratically fitting the fidelity as a function of $\delta h$,
  for $\delta h \lesssim 10^{-6}$.

\bibitem{Vicari-18}
  E. Vicari,
  Decoherence dynamics of qubits coupled to systems at quantum transitions,
  Phys. Rev. A {\bf 98}, 052127 (2018).

\bibitem{NRV-18}
  D. Nigro, D. Rossini, and E. Vicari,
  Scaling properties of work fluctuations after quenches at quantum transitions,
  arXiv:1810.04614 (2018).
  
\bibitem{ZPRS-08} J. Zhang, X. Peng, N. Rajendran, and D. Suter,
  Detection of quantum critical points by a probe qubit,
  Phys. Rev. Lett. {\bf 100}, 100501 (2008).

\bibitem{ZCCL-09} J. Zhang, F. M. Cucchietti, C. M. Chandrashekar,
  M. Laforest, C. A. Ryan, M. Ditty, A. Hubbard, J. K. Gamble, and
  R. Laflamme, Direct observation of quantum criticality in Ising spin
  chains, Phys. Rev. A {\bf 79}, 012305 (2009).

\bibitem{GPSZ-06} T. Gorin, T. Prosen, T. H. Seligman, and M. \v
  Znidari\v c, Dynamics of Loschmidt echoes and fidelity decay,
  Phys. Rep. {\bf 435}, 33 (2006).

\bibitem{JP-09} Ph. Jacquod and C. Petitjean, Decoherence,
  entanglement and irreversibility in quantum dynamical systems with
  few degrees of freedom, Adv. Phys. {\bf 58}, 67 (2009).
  
\bibitem{DGP-10} C. De Grandi, V. Gritsev, and A. Polkovnikov, Quench
  dynamics near a quantum critical point, Phys. Rev. B {\bf 81},
  012303 (2010).

\bibitem{SA-17} M. Serbyn and D. A. Abanin, Loschmidt echo in
  many-body localized phases, Phys. Rev. B {\bf 96}, 014202 (2017).

\bibitem{QSZS-06} H. T. Quan, Z. Song, X. F. Liu, P. Zanardi, and
  C. P. Sun, Decay of Loschmidt echo enhanced by quantum criticality,
  Phys. Rev. Lett. {\bf 96}, 140604 (2006).

\bibitem{RCGMF-07} D. Rossini, T. Calarco, V. Giovannetti,
  S. Montangero, and R. Fazio, Decoherence induced by interacting
  quantum spin baths, Phys. Rev. A {\bf 75}, 032333 (2007).

\bibitem{HGMPM-12} P. Haikka, J. Goold, S. McEndoo, F. Plastina, and
  S. Maniscalco, Non-Markovianity, Loschmidt echo, and criticality: A
  unified picture, Phys. Rev. A {\bf 85}, 060101(R) (2012).

\end{thebibliography}
\end{document}